\documentclass[prb,amsmath,twocolumn,showpacs]{revtex4}
\usepackage{bm}
\usepackage{amssymb}
\usepackage{graphicx}
\def\bx{{\mbox{\boldmath $x$}}}

\def\bn{{\mbox{\boldmath $n$}}}

\begin{document}

\title{Lapse of transmission phase and electron molecules in quantum dots}

\author{S.A. Gurvitz}
\email{shmuel.gurvitz@weizmann.ac.il}

\affiliation{Department of Particle Physics,  Weizmann Institute of
Science, Rehovot 76100, Israel}

\date{\today}


\begin{abstract}
The puzzling behavior of the transition phase through a quantum dot
can be understood in a natural way via formation of the electron
molecule in the quantum dot. In this case, the resonance tunneling
takes place through the quasistationary (doorway) state, which
emerges when the number of electrons occupying the dot reaches a
certain ``critical'' value, $N_{cr}$. Our estimation of this
quantity agrees with the experimental data. The dependence of
$N_{cr}$ on the dot's size is predicted as well.
\end{abstract}

\maketitle

One of the challenging problems in mesoscopic physics is the
puzzling behavior of the transmission phase through a quantum dot,
embedded in an Aharonov-Bohm ring. It was found in series of
experiments performed by the Weizmann group\cite{yac,schus,michal}
that all transmission amplitudes through different resonant levels
of a quantum dot are in phase. This necessarily implies an
unexpected lapse in the evolution of the transmission phase between
different resonant levels. In addition, it was found in recent
measurements \cite{michal} that this phenomenon takes place when the
number of electrons inside the dot reaches a certain ``critical''
value ($N_{cr}\gtrsim 15$) \cite{michal}. In spite of many
publications addressed to these experiments no fully satisfactory
understanding has been found yet \cite{hac2,gef}.

In this Rapid Communication we demonstrate that the observed
phase-lapse behavior of the transmission amplitude can be naturally
explained by implying the formation of electron (Wigner) molecules
inside quantum dots, proposed in recent publications
\cite{reinmann,ghosal,lan1,lan2}. Moreover, this framework allows us
to estimate $N_{cr}$ and then to determine how it is varied with a
size of the dot. In order to explain our model in a proper way, we
first elaborate the physical nature of the transmission phase in the
case of noninteracting and interacting electrons. In particular, we
concentrate on the role of the Pauli principle that prevents
different conductance resonances to carry essentially the {\em same}
internal wave function. This point represents a formidable obstacle
for resolving the puzzling behavior of the transmission phase for
different models of the quantum dot. We demonstrate, however, that
this difficulty can be overcame in the context of the
Wigner-molecule when an unstable state is developed in the middle of
the dot.

\begin{figure}
\includegraphics[width=7cm]{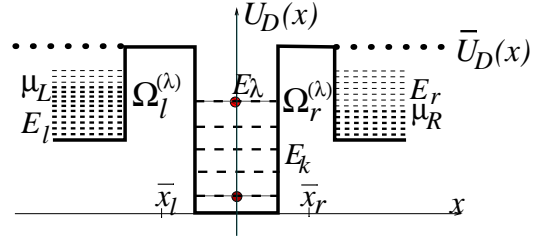}
\caption{(Color online) Resonant tunneling trough a quantum dot.
$\mu_{L(R)}$ are Fermi energies in the left (right) reservoir. The
dotted lines show the potential $\bar U_D(x)$, needed for evaluation
of the bound state wave functions in the Bardeen formula.}
\label{fig1}
\end{figure}

Let us consider the resonant tunneling through a quantum dot,
represented by a potential $U_D(x)$, Fig.~\ref{fig1}. The bottom of
this potential can be moved by the plunger electrode, so that one
observes the current sweeping trough different resonant states
($E_\lambda$) of the dot. We would treat this problem in the
framework of a tunnel Hamiltonian approach. This approach is more
transparent for evaluation of the transmission phase than the
standard scattering theory, in particular, when the Pauli principle
and the electron-electron interaction are taken into account. We
introduce therefore the following tunneling Hamiltonian:
$H=H_L+H_R+H_D+H_T$, where
\begin{align}
&H_{L(R)}=\sum_{l(r)}E_{l(r)}a^\dagger_{l(r)}a_{l(r)}\, ,
  ~~~
  H_D=\sum_kE_kd^\dagger_kd_k+H_C\, ,
  \nonumber\\
&H_T=\Big (\sum_{l,k}\Omega^{(k)}_ld^\dagger_ka_l +l\leftrightarrow
r\Big )+H.c.
  \label{a1}
\end{align}
Here, $a^\dagger_{l,r}(a_{l,r})$ is the creation (annihilation)
operator of an electron in the reservoirs and $d^\dagger_k(d_k)$ is
the same operator for an electron inside the dot. For simplicity, we
consider electrons as spin-less fermions. The term $H_C$ denotes the
Coulomb interaction between electrons in the dot and
$\Omega^{(k)}_l[\Omega^{(k)}_r]$ is the coupling between the states
$E_l(E_r)$ and $E_k$ of the reservoir and the dot, respectively. In
the absence of magnetic field, all couplings $\Omega$ are real.

All parameters of the tunneling Hamiltonian (\ref{a1}) are related
to the initial microscopic description of the system in the
configuration space. For instance, the coupling
$\Omega^{(k)}_{l(r)}$ is given by the Bardeen formula\cite{bar}
\begin{align}
\Omega^{(k)}_{l(r)}=-{\hbar^2\over
2m}\int_{\bx\in\Sigma_{l(r)}}\phi_k(\bx )
\stackrel{\leftrightarrow}\nabla_{\bn}\chi_{l(r)}(\bx )d\sigma\ ,
\label{a2}
\end{align}
where $\phi_k(\bx )$ and $\chi_{l(r)}(\bx )$ are the electron wave
functions inside the dot and the reservoir, respectively, and
$\Sigma_{l(r)}$ is a surface inside the left (right) barrier that
separates the dot from the corresponding reservoir. It is important
to point out that $\phi_k(\bx )$ in Eq.~(\ref{a2}) is a {\em bound
state} wave function for the ``inner'' potential. The latter
coincides with the original potential inside the surface $\Sigma$
and a constant outside this region. On the other hand,
$\chi_{l(r)}(\bx )$ is a {\em non-resonant} scattering wave function
in the ``outer'' potential, which coincides with the original
potential outside the surface $\Sigma$ and a constant inside this
region \cite{g3}.

In one-dimensional case (Fig.~\ref{fig1}), the separation surface
$\Sigma$ becomes the separation point, $\bar x$, inside the barrier,
Fig.~\ref{fig1}. Then Eq.~(\ref{a2}) can be rewritten as \cite{g1}
\begin{align}
\Omega^{(k)}_{l(r)}=-(\kappa_k/m)\phi_k(\bar x_{l(r)})\chi_{l(r)}
(\bar x_{l(r)})\, , \label{a3}
\end{align}
where $\kappa_k=\sqrt{2m[U_D(\bar x_{l,r})-E_k]}$ and $\phi_k(x)$ is
the bound state wave function in the potential $\bar U_D(x)$
(Fig.~\ref{fig1}). The separation points $\bar x_{l,r}$ are to be
taken inside the left (right) barrier as indicated in
Fig.~\ref{fig1} and far away from the classical turning points
\cite{g2}.

We start with non-interacting electrons, $H_C=0$ in Eq.~(\ref{a1}).
Then the electron transport through the level $E_\lambda$ can be
described by the time-dependent Schr\"odinger equation
$i\hbar\partial_t|\Psi(t)\rangle=H|\Psi(t)\rangle$ for a single
electron. Taking the stationary limit we obtain the Landauer formula
for the total current, with the transmission amplitude given by the
Bright-Wigner formula
\begin{align}
t_\lambda(E)={\cal N}{\Omega^{(\lambda)}_L\Omega^{(\lambda)}_R \over
E-E_\lambda +i(\Gamma^{(\lambda)}_L+\Gamma^{(\lambda)}_R)/2}\, ,
\label{a4}
\end{align}
where ${\cal N}=-2\pi(\varrho_L\varrho_R)^{1/2}$ and
$\Gamma^{(\lambda)}_{L,R} =2\pi (\Omega_{L,R}^{(\lambda
)})^2\varrho_{L,R}$ are the partial widths, and $\varrho_{L(R)}$ is
the density of states in the left (right) reservoir. We assumed that
$\Omega_{l,r}^{(\lambda )}\equiv \Omega_{L,R}^{(\lambda )}$ are
weakly dependent on $E_{l,r}$.

The corresponding evolution of the resonance transmission phase for
different states $|\lambda\rangle$ is determined by sign of the
product of $\Omega^{(\lambda )}_L\Omega^{(\lambda )}_R$. Since the
reservoir states $\chi_{l,r}$ are not affected by the plunger
voltage, one finds from Eq.~(\ref{a3}) that the evolution of the
${\mbox{sign}}\,[\Omega^{(\lambda )}_L\Omega^{(\lambda )}_R]$ is
given by the sign of the product $\phi_\lambda (\bar
x_l)\phi_\lambda (\bar x_r)$. The latter is positive or negative,
depending on the number of nodes of $\phi_\lambda (x)$. Hence, it is
clear that the non-interacting electron model cannot explain the
same sign for all resonances, observed in Ref.\cite{schus} (see also
Refs. \cite{hac2,gef1}).

Consider $N$ {\em interacting} electrons trapped inside the dot.
Despite the electron-electron interaction, the coupling amplitudes
$\Omega_{L,R}$ can still be evaluated by using the same
multi-dimensional overlapping formula (\ref{a2}), as in the
non-interacting case. Indeed, the many-body tunneling can be
considered as one-body tunneling, but in the many-dimensional space.
In this case, the wave-function $\chi_{l(r)}(\bx )$ is replaced by
$\chi_{l(r)}(x_{N+1})\Phi_N^{(0)}(x_1,\ldots ,x_N)$, where
$\chi_{l(r)}$ is the wave function of tunneling electron in the left
(right) reservoir and $\Phi_N^{(0)}$ is the ground state wave
function of $N$ electrons inside the dot. The wave-function
$\phi_k(\bx )$ corresponds to $\Phi_{N+1}^{(0)}(x_1,\ldots
,x_{N+1})$, which is the lowest energy state (ground state) of $N+1$
electrons in the inner potential of the dot ($\bar U_D$ in
Fig.~\ref{fig1}).

Taking $\bn$ along a coordinate of the tunneling electron,
$x_{N+1}$, we can integrate over $x_1,\ldots ,x_N$ in Eq.~(\ref{a2})
thus reducing this equation to Eq.~(\ref{a3}) with $\phi_n$ being
replaced by the overlap function
\begin{align}
\varphi_N(x_{N+1} )=\langle x_{N+1},\Phi_N^{(0)}
|\Phi_{N+1}^{(0)}\rangle\, . \label{aa4}
\end{align}
Therefore, the sign of $\Omega^{(\lambda)}_L\Omega^{(\lambda)}_R$ is
determined by the sign of $\varphi_N(\bar x_l)\varphi_N(\bar x_r)$.

By applying the mean-field approximation, we can write
$|\Phi_N^{(0)}\rangle$ and $|\Phi_{N+1}^{(0)}\rangle$ as a product
of one-electron states (orbitals) in the effective single-particle
potential, $\bar U_D+U_C$, where $\bar U_D$ is the inner part of
quantum-dot potential (Fig.~\ref{fig1}) and $U_C(x)$ is the
mean-field describing the electron-electron interaction. As a
result, the overlap function $\varphi_N(x)$ is a bound state wave
function in the potential $\bar U_D(x)+U_C(x)$, corresponding to one
of the orbitals. Since the lowest energy state is always
nodeless\cite{sim}, one might assume that $\varphi_N(x)$ is also a
nodeless one, so that the sign of $\varphi_N(\bar x_l)\varphi_N(\bar
x_r)$ would be the same sign for all resonances. This, however, is
not correct because of the Pauli principle. Indeed, due to the
anti-symmetrization, any two orbitals in the product of the wave
functions representing $|\Phi_{N+1}^{(0)}\rangle$ cannot be the
same. Since the lowest state is already occupied, the wave function
$\varphi_N(x_{N+1} )$ must correspond to a higher non-occupied
orbital, and therefore it cannot be nodeless. Hence, the Pauli
principle would create serious problems in any attempt to explain
the same sign for all resonances \cite{schus} in a framework of the
mean-field description of the electron-electron interaction.

Note that this problem cannot be resolved even by assuming large
coupling with reservoirs, so that the resonances are overlap.
Indeed, the problem is related only to the {\em inner} component of
the resonant state, Eqs.~(\ref{a2}) and (\ref{a3}). The latter is
eventually brought by the plunger below the Fermi level, $\mu_R$,
blocking an appearance of the resonance above the Fermi level with a
similar inner component.

The same situation holds in a more general case, when the
interaction term $U_C$ varies with each new electron trapped inside
the dot, $U_C\to U^{(N)}_C$ (Koopman`s theorem is violated). One
finds that due to the central symmetry of the self-consistent
potential such a variation of $U_C$ with $N$ would not affect the
number of nodes in the overlap function $\varphi_N(x_{N+1} )$. As a
result, the sign of the transmission amplitude would fluctuate
between $\pm 1$ for different resonances.

We illustrate this point by evaluating the overlap function
$\varphi_N(x_{N+1} )$, Eq.~(\ref{aa4}), for $N=0$ and $N=1$.  In the
first case, $\varphi_0(x_1)$ coincides with the wave function of the
lowest energy state, $\Phi_1^{(0)}(x_1)\equiv\tilde\phi_0(x_1)$, in
the inner potential $\bar U_D$, Fig.~\ref{fig1}. This wave function
is nodeless. The second overlap function is $\varphi_1(x_2 )=\langle
x_2,\Phi_1^{(0)} |\Phi_2^{(0)}\rangle$, where
$\Phi_2^{(0)}(x_1,x_2)=[\phi_0(x_1)\phi_1(x_2)-\phi_0(x_2)\phi_1(x_1)]/\sqrt{2}$
is the lowest energy state of two electrons in the potential $\bar
U_D+U_C^{(2)}$. Here $\phi_{0,1}$ represent the two first orbitals
in this potential.  One easily finds that
\begin{align}
\varphi_1(x_2)=\int\Phi_1^{(0)}(x_1)\Phi_2^{(0)}(x_1,x_2)dx_1
=c_0\phi_1(x_2)\, , \label{a5}
\end{align}
where $c_0=\int\tilde\phi_0(x_1)\phi_0(x_1)dx_1/\sqrt{2}$. (The
second term is zero, since $\tilde\phi_0$ and $\phi_1$ are
orthogonal due to the opposite parities). Therefore, $\varphi_1$
contains one node, so that the corresponding transition amplitude
changes its sign.

The same behavior of the overlap function would persist for any $N$.
For instance, one easily obtains for $N=2$ that
$\varphi_2(x_3)\propto c_0\phi_3(x_3 )-c_{13}\phi_1(x_3 )$, where
the coefficients $c_0=\langle\tilde\phi_1|\phi_1\rangle$,
$c_{13}=\langle\tilde\phi_0|\phi_2\rangle$ and $\tilde\phi$, $\phi$
are the orbitals in the potentials, $\bar U_D(x )+U_C^{(1)}(x)$ and
$\bar U_D(x)+U_C^{(2)}(x)$, respectively. Since $c_{13}\ll c_0$, the
overlap function $\varphi_2$ would contain an additional node in a
comparison to $\varphi_1$. Thus, by assuming the $N$ dependence of
the mean-field effective potential, we are still not able to explain
the puzzling behavior of the transmission phase.

The above consideration was based on symmetry arguments applied to
electrons moving in a spherically symmetric mean-field central
potential. In fact, the central mean-field picture for
two-dimensional quantum dots was challenged in recent publications
\cite{reinmann,lan1,ghosal,lan2}. It was suggested that due to the
strong inter-electron repulsion inside the dot, spontaneous symmetry
breaking takes place leading to the formation of electron molecules.
As a result, the electrons appear on the ring (rings) around the
dot's center. This idea was substantiated by unrestricted
Hartree-Fock calculations or by using other computational
techniques\cite{reinmann,lan1,ghosal,lan2}.

In principle, if the symmetry is broken, the overlap function
$\varphi_N(x_{N+1} )$, Eq.~(\ref{aa4}), could be very different from
the corresponding orbital $\phi_N(x_{N+1} )$ in the spherical
symmetric potential. Therefore, it is desirable to investigate the
evolution of the transmission phase in this case. Consider again the
overlap function $\varphi_1(x_2 )=\langle\bx_2,\Phi_1^{(0)}
|\Phi_2^{(0)}\rangle$, but now without the mean-field approximation,
as in Eq.~(\ref{a5}). In fact, by taking the parabolic confining
potential, the two-electron wave function $|\Phi_2^{(0)}\rangle$ can
be exactly calculated\cite{wagner}, since relative and
center-of-mass coordinates of two electrons are decoupled in the
total Hamiltonian. As a result,
$\Phi_2^{(0)}(x_1,x_2)=\phi_{cm}(x_1+x_2)\phi_r(x_2-x_1)$, where
$\phi_r(-x)=-\phi_r(x)$ due to the Pauli principle. Such a wave
function peaks for $x_1=-x_2$ and therefore it would bear the
features of a two-electron molecule\cite{lan2}. One finds from
Eq.~(\ref{aa4}),
\begin{align}
\varphi_1(x_2)=\int\tilde\phi_0(x_1)\phi_{cm}(x_1+x_2)\phi_r(x_2-x_1)
dx_1\, . \label{a6}
\end{align}
Taking into account that the values of $x_1$ which mainly contribute
to the integral (\ref{a6}) are localized inside the dot and that the
wave function $\phi_r(x)$ is the odd one, we find that the overlap
function changes its sign when the argument varies from $\bar x_l$
to $\bar x_r$, Fig.~\ref{fig1}. Hence, $\varphi_1$ displays one
node, as in the spherically symmetric mean-field potential,
Eq.~(\ref{a5}).

One can continue with the same arguments for the three and more
electron molecules, where the electrons are placed on the ring. The
corresponding ground state wave functions $|\Phi_N^{(0)}\rangle$
would represent a fully anti-symmetrized product of the original
(site) {\em nodeless} orbitals \cite{lan1,ghosal}. Yet, the overlap
function Eq.~(\ref{aa4}) cannot be nodeless. As a result, the sign
of $\varphi_N(\bar x_l)\varphi_N(\bar x_r)$ would fluctuate with
$N$. One can demonstrate it rather easily for $N=3,4$. Although it
would be hard to extend such a demonstration for large $N$, there is
no reason to expect that the sign of $\varphi_N(\bar
x_l)\varphi_N(\bar x_r)$ ceases to fluctuate when $N$ increases.

\begin{figure}
\includegraphics[width=8cm]{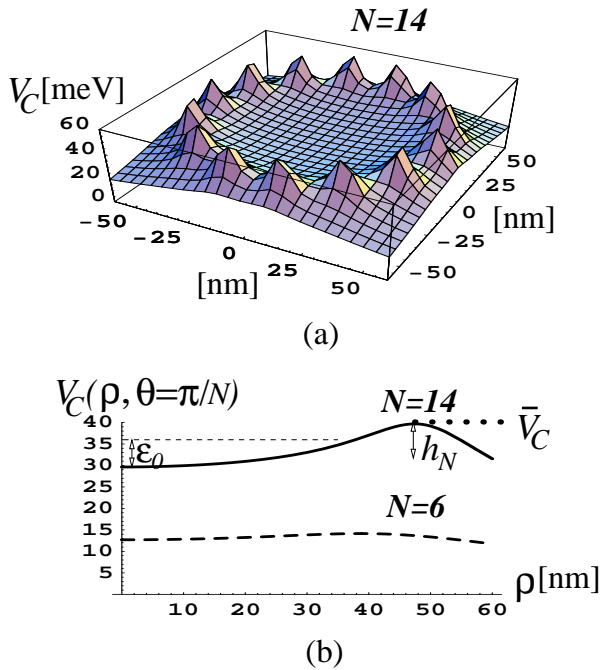}
\caption{(Color online) (a) Electrostatic trap generated by 14
electrons placed on the ring of the radius 50 nm. (b) The radial
profile of the Coulomb potential along the potential valley.
$\varepsilon_0$ is the ground state energy in the potential $\bar
V_C$, representing the inner part of the trap $V_C$.} \label{fig2}
\end{figure}

It seems from the above arguments that the rotational symmetry
breaking (the electron-molecule formation) cannot explain the
evolution of the transmission phase observed in the experiments
\cite{schus,michal}. Nevertheless, there is an additional feature of
the electron molecule, which has not been yet utilized. That is due
to the electrons located on the ring (rings) would develop an
additional (inner) electrostatic trap inside the dot when their
number ($N$) is large enough. As an example, we display in
Fig.~\ref{fig2}a such a potential, $V_C(\bx )=\sum_{j}^N
e^2/\epsilon |\bx-\bx_j|$, produced by $14$ electrons equally
distributed on the ring, where $\epsilon =13.6$ is the dielectric
constant of the medium. The radius of the ring ($R=50$ nm) is taken
close to the dot's size in Ref. \cite{michal}. The radial profile of
this potential along the angle $\theta =\pi/N$, where the potential
height on the ring is minimal, is shown in Fig.~\ref{fig2}b for two
values of $N$. It appears that the trap is not well developed for
$N=6$, but it is already pronounced for $N=14$.

A minimum number of electrons in the dot sufficient to develop the
trap with one bound state inside it can be estimated from the
condition that the barrier height, $h_N$ in the Fig.~\ref{fig2}b,
reaches the ground state energy $\varepsilon_0$. We estimate the
latter as $\pi^2\hbar^2/m^*R^2$, where $m^*$ is the effective
electron mass ($m^*/m_0=0.067$). For instance, one finds
$\varepsilon_0=4.5$ meV for $R=50$ nm. Then the condition
$h_N=\epsilon_0$ corresponds to $N\simeq 10$, which is a minimal
(``critical'') number of electrons, $N_{cr}$, enabled to hold a
resonance state. This value is an approximate agreement with that
found in \cite{michal}. In fact, a more elaborate, semi-classical
estimations of $N_{cr}$ approximately produce the same number
[$h_N\simeq 10$ meV for $N=14$, Fig.~\ref{fig2}b]\cite {fn}.

The state $|\varepsilon_0\rangle$ in the inner part of the trap,
$\bar V_C$, Fig.~\ref{fig2}b, is not stable due to the symmetry
breaking, leading to formation of the electron molecule.
Nevertheless, this state is important in formation of the
($N+1$)-electron molecule by adding an additional electron to the
$N$-electron system. Indeed, one expects that the overlap function
(\ref{aa4}) for the electron states on the ring is suppressed in
comparison to the same overlap for the central mean-field potential.
The reason is that all electrons are shifted from their positions
whenever an additional electron is placed on the ring. This is in
contrast to the mean-field description, where the $N$-electron core
is not modified. On the other hand, if the electron is placed in the
center of the ring, it distorts the remaining $N$ electrons in a
minimal way. We expect therefore that the corresponding overlap
function is large, as in the case of the central mean-field
potential. Hence, such an unstable state $|\varepsilon_0\rangle$ in
the middle of the dot would play a role of a ``doorway'' state in
formation of the ($N+1$)-electron molecule.

It follows from the same arguments that the electron transport would
proceed through such an unstable state when the quantum dot coupled
with the reservoirs. Since this doorway state is of the lowest
energy in the inner trap, $\bar V_C$ (Fig.~\ref{fig2}), it is
nodeless. The crucial point here is that this state is eventually
not occupied, when it is brought by the plunger below the Fermi
levels of the reservoirs. Indeed, it is not turned to a stable state
below the Fermi levels due to the symmetry breaking, but it always
decays to the ring states. Therefore, this state is never blocked by
the Pauli principle to carry the resonant transport through it, when
it is above the Fermi level $\mu_R$, Fig.~\ref{fig1}. As a result,
all transmission amplitudes for any $N>N_{cr}$ would be in phase.

In fact, by taking into account the electron spin, one finds that
two electrons with the same spatial (nodeless) wave functions are
allowed to occupy the lowest energy states. Therefore, even if the
state $|\varepsilon_0\rangle$ in the center of the dot becomes a
stable one for some values of $N$, the resonant transport would
proceed through an unstable state of the two electrons (with
opposite spin) inside the dot. The corresponding overlap function
would be again nodeless.

Note that although the doorway-state energy is the lowest one for
the inner trap, $\bar V_C$, it exceeds the energy of the ring
states. Therefore, the ring states would appear inside the bias
voltage before the doorway state. We can assume, however, that the
ring states are not well separated in the energy from the doorway
state, which dominates the resonant current. It was also taken into
account that in the presence of the Coulomb interaction, the shift
of the resonance energy due to tunneling is different for different
levels \cite{imry}. In particular, the broad resonance is shifted
down more than the narrow one \cite{imry}. As a result the doorway
state could have a lower energy than the ring states.

\begin{figure}
\includegraphics[width=7cm]{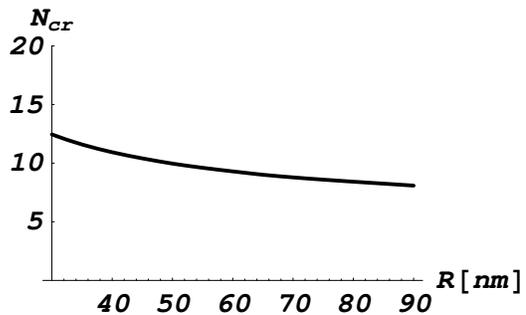}
\caption{Dependence of the ``critical'' number of electrons on the
dot's radius.} \label{fig3}
\end{figure}

One of the consequences of our model is an existence of the critical
number of electrons in the dot, which is necessary for formation of
the resonant state inside the dot ($N_{cr}$).  This number would
vary with the dot's size. Such a dependence of $N_{cr}$ on the
radius of the dot ($R$), obtained from our estimation,
$h_N=\varepsilon_0$, is shown in Fig.~\ref{fig3}. One finds from
this figure that this dependence is rather weak. The critical number
slightly decreases with an increase of the dot's size.

In summary, we demonstrated that the unusual behavior of the
resonant phase, observed in interference experiments, can be
considered as a strong evidence for formation of electron molecules
in quantum dots. This structure would produce an electrostatic trap,
containing an unstable (doorway) state localized in the center of
the dot, whenever the number of electrons occupying the dot is large
enough, $N>N_{cr}$. Then such an unstable state would carry the
electron transport through the dot irrespective of the value of $N$.
This would appear as if the different transmission amplitudes are in
phase. Our prediction for the dependence of $N_{cr}$ on the dot's
radius can be experimentally verified.


\begin{thebibliography}{999}
\bibitem{yac} A. Yacoby, M. Heiblum, D. Mahalu, and H. Shtrikman,
Phys. Rev. Lett. {\bf 74}, 4047 (1995).
\bibitem{schus} R. Schuster, E. Buks, M. Heiblum,
D. Mahalu, V. Umansky and H. Shtrikman, Nature {\bf 385}, 417
(1997).
\bibitem{michal} M. Avinun-Kalish, M. Heiblum, O. Zarchin, D. Mahalu and V. Umansky,
Nature, {\bf 436}, 529 (2005).
\bibitem{hac2} G. Hackenbroich, Phys. Rep. {\bf 343}, 463 (2001).
\bibitem{gef} C. Karrasch, T. Hecht, A. Weichselbaum, Y. Oreg, J. vonDelft
and V. Meden, Phys. Rev. Lett. {\bf 98}, 186802 (2007).
\bibitem{reinmann} S. Reimann and M. Manninen, Rev. Mod. Phys. {\bf 74},
1283 (2002).
\bibitem{lan1} C. Yannouleas and U. Landman, Phys. Rev. B {\bf 68},
035325 (2003).
\bibitem{ghosal} A. Ghosal, A.D. G\"{u}\c{c}l\"{u}, C.J. Umrigar, D.
Ullmo and  H.U. Baranger, Nature Phys. {\bf 2}, 336(2006).
\bibitem{lan2} C. Yannouleas and U. Landman,
Rep. Prog. Phys. {\bf 70}, 2067 (2007).
\bibitem{bar} J. Bardeen, Phys. Rev. Lett. {\bf 6}, 57 (1961).
\bibitem{g3} S.A. Gurvitz, in {\em Multiple facets of quantization and
supersymmetry, Michael Marinov Memorial Volume},  p. 91 (World
Scientific, 2002).
\bibitem{g1} S.A. Gurvitz, Phys. Rev. A {\bf 38}, 1747 (1988).
\bibitem{g2} S.A. Gurvitz, P.B. Semmes, W. Nazarewicz, and T. Vertse,
T, Phys. Rev. {\bf A69}, 042705 (2004).
\bibitem{gef1} A. Silva, Y. Oreg, and Y. Gefen, Phys. Rev.
{\bf B66}, 195316 (2002).
\bibitem{sim} M. Reed and B. Simon, {\em Methods of Modern
Mathematical Physics}, vol. IV (Academic Press, New York, 1977).
\bibitem{wagner} M. Wagner {\em et al.}, Phys. Rev. B {\bf 45}, 1951
(1992).
\bibitem{fn} In fact, a two-ring structure is expected for $N=14$
electrons \cite{reinmann,lan2}. Although this creates a more
complicated trap, our simple estimations of $N_{cr}$, based on the
effective one-dimentional potential, Fig. 2(b), remain the same. For
such estimations it is sufficient to consider all electrons on one
ring of the average radius of two rings. Note that $N_{cr}$ is
weakly dependent on the ring's radius, Fig. 3.
\bibitem{imry} P.G. Silvestrov and Y. Imry, Phys. Rev. Lett. {\bf
85}, 2565 (2000); {\em ibid}, New J. Phys. {\bf 9}, 125 (2007).
\end{thebibliography}
\end{document}